\documentclass[a4paper]{report}
\usepackage[utf8]{inputenc}
\usepackage[T1]{fontenc}
\usepackage{RJournal}
\usepackage{amsmath,amssymb,array}
\usepackage{booktabs}
\usepackage{thumbpdf,lmodern}
\usepackage{amsmath}
\usepackage{framed}
\usepackage{bbm}
\usepackage{algorithm}
\usepackage[noend]{algpseudocode}
\usepackage{float}
\usepackage[caption = false]{subfig}
\usepackage{mathtools}
\newcommand{\verteq}{\rotatebox{90}{$=$}}

\begin{document}

\sectionhead{Contributed research article}
\volume{XX}
\volnumber{YY}
\year{20ZZ}
\month{AAAA}

\begin{article}
\title{The mbsts package: Multivariate Bayesian Structural Time Series Models in R}
\author{by Ning Ning and Jinwen Qiu}

\maketitle

\abstract{
The multivariate Bayesian structural time series (MBSTS) model \citep{qiu2018multivariate} is a general machine learning model that deals with inference and prediction for multiple correlated time series, where one also has the choice of using a different candidate pool of contemporaneous predictors for each target series. The MBSTS model has wide applications and is ideal for feature selection, time series forecasting, nowcasting, inferring causal impact, and others. This paper demonstrates how to use the R package \pkg{mbsts} for MBSTS modeling, establishing a bridge between user-friendly and developer-friendly functions in the package and the corresponding methodology. Object-oriented functions in the package are explained in the way
that enables users to flexibly add or deduct some components, as well as to simplify or complicate some settings. 
}

\section{Introduction}
Structural time series models are state space models for time series data. They are constructed in terms of components each of which has a direct interpretation. For example, one may consider a decomposition in which a series can be seen as the sum of trend and regression components.
The multivariate Bayesian structural time series (MBSTS) model \citep{qiu2018multivariate} is a generalized version of many structural time series models and is constructed as the sum of a trend component, a seasonal component, a cycle component, a regression component, and an error term, where each component provides an independent and additional effect. Users have flexibility in choosing these components and are free to construct their specific forms, for example adding on a regression component with predictors generated through data mining as that in \citep{Jammalamadaka2019Predicting}. 
The MBSTS model uses the Bayes selection technique via Markov chain Monte Carlo (MCMC) methods to select among a set of contemporary predictors, thus one does not need to commit to a fixed set of predictors. Specifically, the variable selection technique uses a ``spike and slab'' approach, through which a different set of predictors can be selected in each MCMC iteration. Then important predictors will be selected according to their overall frequency of numbers being selected over the total number of MCMC iterations. The multivariate structure and the Bayesian framework allow the model to take advantage of the association structure among target series. 

The MBSTS model and its univariate counterpart, the BSTS model \citep{scott2014predicting,scott2015bayesian}, have wide applications in causal inference (see, e.g., \cite{brodersen2015inferring}), heath care (see, e.g., \cite{kurz2019effect}), spatial analysis (see, e.g., \cite{qiu2019evolution}), artificial intelligence (see, e.g., \cite{Jammalamadaka2019Predicting}), cryptocurrency (see, e.g., \cite{jalan2019if}), medicine (see, e.g., \cite{talaei2019predictive}), airline industry (see, e.g., \cite{talaei2019predictive}), environmental science (see, e.g., \cite{droste2018decentralization}), renewable energy  (see, e.g., \cite{jiang2013very}), political analysis (see, e.g., \cite{xu2017generalized}), social media  (see, e.g., \cite{welbers2018social}), and etc. Upon demands from users of the BSTS/MBSTS model in understanding the methodology and establishing a user-friendly interface with sufficient flexibility, the R package \pkg{mbsts} \citep{qiu2021package} is developed, which is available on CRAN at  \href{https://cran.r-project.org/web/packages/mbsts/}{https://cran.r-project.org/web/packages/mbsts/}. Our contributions are four-fold:
\begin{enumerate}
\item \textbf{Functional contribution.} The \pkg{mbsts} package is the only package that implements the MBSTS model \citep{qiu2018multivariate}. Therefore, it is one that uses Bayesian tools for model fitting, prediction, and feature selection on multivariate correlated time series data, where different contemporaneous predictors could be selected for different target series. The \pkg{bsts} package \citep{scott2021package} implements the BSTS model which only works on one-dimensional time series data.

\item \textbf{User-friendly interface.} The \pkg{mbsts} package is developed with the purpose to give users a nice and easy experience. 
Users of the \pkg{mbsts} package can conduct model training using the \code{mbsts\_function} function, perform prediction using the \code{mbsts.forecast} function, retrieve parameter estimation results using the \code{para.est} function, visualize feature selection results using the \code{plot\_prob} function,  diagnose convergence using the \code{plot\_cvg} function,  and visualize time series components training results using the \code{plot\_comp} function. Different from some machine learning algorithms that have to resort to cloud computing such as \cite{ning2021iterated} and \cite{ionides2022iterated}, the \pkg{mbsts} package can perform well on personal computers in non-trivial case studies.

\item \textbf{Technical contribution.} Different from the \pkg{bsts} package using R and C/C++ for univariate time series analysis, the \pkg{mbsts} package fully relies on  R's classes and methods system in a concise way for multivariate time series. 
The \pkg{bsts} package uses S$3$ classes and methods while the \pkg{mbsts} package uses the more formal and advanced S$4$ classes and methods. The \pkg{mbsts} package is developed as efficiently as possible, where
feature selection and model training can be conducted at the same time using the main function \code{mbsts\_function}.
 
\item \textbf{R developer-friendly structure.} The \pkg{mbsts} package aims to help R developers easily understand and edit. Therefore, it is also developed as concisely as possible with a simple structure. It is editable to users that only understand R, while the \pkg{bsts} package has a much more complicated structure and R developers have to understand well of C/C++ in order to edit the \pkg{bsts} package. In this paper, we start with the illustration of the model structure, such that R developers and package users only need to read this paper to fully understand the model and use the package.
\end{enumerate}

\section{The MBSTS model and the \code{sim\_data} function} 
\label{sec:models}
In this section, we introduce the MBSTS model in terms of its basic model structure and flexible state components. The MBSTS model is further illustrated using the \code{sim\_data} function provided in the \textbf{mbsts} package, where data reproducibility is reached by setting the seed:
\begin{example}
R>  set.seed(1)
\end{example}

\subsection{Model structure}
The MBSTS model is constructed as the sum of components and one can flexibly select suitable components:
\begin{equation} \label{eq:st}
\tilde{y}_t=\tilde{\mu}_t+\tilde{\tau}_t+\tilde{\omega}_t+\tilde{\xi}_t+\tilde{\epsilon}_t,     \ \ \ \ \ t=1,2,\cdots,n,
\end{equation}
where  $\tilde{y}_t,\ \tilde{\mu}_t,\ \tilde{\tau}_t,\ \tilde{\omega}_t,\ \tilde{\xi}_t$,
and $\tilde{\epsilon}_t$ are $m$-dimensional vectors, representing target time series, a linear trend component, a seasonal component, a cyclical component, a regression component, and an observation error term, respectively. 
Let $\tilde{y}_t=[y_t^{(1)},\cdots,y_t^{(m)}]^T$ and let the $m$-dimensional vector $\tilde{\epsilon}_t$  distributed as
$\tilde{\epsilon}_t\stackrel{iid}\sim \  N_m(0,\Sigma_\epsilon)$ where $\Sigma_\epsilon$ is a $m\times m$-dimensional positive definite matrix. The simulated dataset \code{data} will be constructed merely for illustration purpose: there are $m=2$ target time series each of which has $505$ observations, the variance of the first  target series $y_t^{(1)}$ is $1.1$, the variance of the second target series $y_t^{(2)}$ is $0.9$, and the covariance between $y_t^{(1)}$ and $y_t^{(2)}$ is $0.7$. Interested readers are referred to \citep{qiu2018multivariate}
for MBSTS models with more correlated target times series, as well as extensive model performance examinations over different scenarios.
\begin{example}
R> n<-505 #n: sample size
R> m<-2 #m: dimension of target series 
R> 	
R> #covariance matrix of target series
R> cov<-matrix(c(1.1,0.7,0.7,0.9), nrow=2, ncol=2) 
\end{example}

Regression analysis is a statistical approach for estimating the relationships between dependent variables and independent variables which are also called predictors, covariates, or features. The regression component $\tilde{\xi}_t=[\xi_{t}^{(1)},\cdots,\xi_{t}^{(m)}]^T$ is written as follows:
\begin{equation} \label{eq:regression}
\xi^{(i)}_t=\beta_i^Tx^{(i)}_t.
\end{equation}
where  $x_t^{(i)}=[x_{t1}^{(i)},\dots,x_{tk_i}^{(i)}]^T$ is the pool of all available predictors at time $t$ for target series $y^{(i)}$, and $\beta_i=[\beta_{i1},\dots,\beta_{ij},\dots,\beta_{ik_i}]^T$ represents corresponding static regression coefficients. Recall that $\tilde{y}_t=[y_t^{(1)},\cdots,y_t^{(m)}]^T$ and let
$B=[\beta_1,\dots,\beta_i,\dots, \beta_m]$ denote a $k\times m$-dimensional matrix. Then we can rewrite $\tilde{\xi}_t$ as  
\begin{equation} 
\tilde{\xi}_t=\operatorname{diag}(B^T\tilde{x}_t).
\end{equation}
In \code{data}, the regression component  is generated by setting $B$ and $\tilde{x}_t$ as follows:
\begin{equation}
\begin{gathered}
B=\begin{bmatrix}
2 & 0 & 2.5 & 0 & 1.5 & -2 & 0 & 3.5\\
-1.5 & 4 & 0 & 2.5  & -1 & 0 & -3 & 0.5
\end{bmatrix}^T, \\
\tilde{x}_t=\begin{bmatrix}
x_{t1} & x_{t2} & x_{t3}&  x_{t4}& x_{t5} & x_{t6}& x_{t7} & x_{t8}\\
x_{t1} & x_{t2} & x_{t3}&  x_{t4}& x_{t5} & x_{t6}& x_{t7} & x_{t8}
\end{bmatrix}^T,\\
x_{t1}\stackrel{iid}\sim N(5,5^2),\ \ \  x_{t2}\stackrel{iid}\sim Pois(10),\ \ \ x_{t3}\stackrel{iid}\sim Pois(5),\ \ \  x_{t4}\stackrel{iid}\sim N(-2,5),\\
x_{t5}\stackrel{iid}\sim N(-5,5^2),\ \ \  x_{t6}\stackrel{iid}\sim Pois(15),\ \ \  x_{t7}\stackrel{iid}\sim Pois(20),\ \ \ x_{t8}\stackrel{iid}\sim N(0,10^2).
\end{gathered}
\end{equation}
The coding realization for the regression component is given by
\begin{example}
R> #Regression component
R> #coefficients for predictors
R> beta<-t(matrix(c(2,-1.5,0,4,2.5,0,0,2.5,1.5,-1,-2,0,0,-3,3.5,0.5),nrow=2,ncol=8)) 
R> 
R> #predictors
R> X1<-rnorm(n,5,5^2)
R> X4<-rnorm(n,-2,5)
R> X5<-rnorm(n,-5,5^2)
R> X8<-rnorm(n,0,100)
R> X2<-rpois(n, 10)
R> X6<-rpois(n, 15)
R> X7<-rpois(n, 20)
R> X3<-rpois(n, 5)
R> X<-cbind(X1,X2,X3,X4,X5,X6,X7,X8)
\end{example}

\subsection{Trend component}
A trend is the long-term growth of time series, and it can be further decomposed into two components: level and slope. Level represents the actual mean value of the trend and slope represents the tendency to grow or decline from the trend.
The trend component $\tilde{\mu}_t$   is generated by a generalization of the local linear trend model where the slope exhibits stationarity in the form as:
\begin{align} 
	\tilde{\mu}_{t+1}&=\tilde{\mu}_t+\tilde{\delta}_t+\tilde{u}_t, \ \quad\quad\quad \  \ \;\; \tilde{u}_t\stackrel{iid}\sim \  N_m(0,\Sigma_\mu),\label{eq:trend}\\
	\tilde{\delta}_{t+1}&=\tilde{D}+\tilde{\rho}(\tilde{\delta}_t-\tilde{D})+\tilde{v}_t, \ \ \  \ \  \tilde{v}_t\stackrel{iid}\sim \  N_m(0,\Sigma_\delta).\label{eq:slope}
\end{align}
Here, $\tilde{\delta}_t$ is a $m$-dimensional vector as short-term slope and $\tilde{D}$ is a $m$-dimensional vector as the long-term slope or in another name as level, which enables the model to incorporate short-term information with long-term information. $\tilde{\rho}$
is a $m\times m$-dimensional diagonal matrix with diagonal entries $0\le \rho_{ii}\le1$ for $i=1,2,\cdots,m$, to represent the learning rates at which the local trend is updated for $\{y_t^{(i)}\}_{i=1,2,\cdots,m}$. 
In the trend component, covariance matrices for error terms $\Sigma_\mu$ and $\Sigma_\delta$ are assumed to be $m\times m$-dimensional diagonal matrices. 


Seasonality is a characteristic of a time series in which the data has regular and predictable changes that recur every period.
The seasonal component $\tilde{\tau}_t=[\tau^{(1)}_t,\cdots,\tau^{(m)}_t]^T$  is generated as follows:
\begin{equation} \label{eq:seasonal}
	\tau_{t+1}^{(i)}=-\sum_{k=0}^{S_i-2}\tau^{(i)}_{t-k}+w^{(i)}_t,  \ \ \  \ \ \tilde{w}_t=[w^{(1)}_t,\cdots,w^{(m)}_t]^T\stackrel{iid}\sim  \ N_m(0,\Sigma_\tau),
\end{equation}
where $S_i$ represents the number of seasons for $y^{(i)}$ and $\tilde{\tau}_t$ is a $m$-dimensional vector
denoting their joint contribution to the observed target time series $\tilde{y}_t$.
The model allows for various seasonal components with different periods for each target series $y^{(i)}$, such as one can include a seasonal component with $S_i=7$ to capture the day-of-the-week effect for one target series, and $S_j=30$  to capture the day-of-the-month effect for another target series. 
In the seasonal component, the covariance matrix for the error term $\Sigma_\tau$ is assumed to be a $m\times m$-dimensional diagonal matrix.


The cyclical effect refers to regular or periodic fluctuations around the trend, revealing a succession of phases of expansion and contraction. In contrast to seasonality that is always of fixed and known periods, a cyclic pattern exists when data exhibits ups and downs that are not of fixed periods.
The cycle component $\tilde{\omega}_t$ is formulated as a dynamic in a fully-coupled dynamical system:
\begin{equation} \label{eq:cycle}
	\begin{gathered}
		\tilde{\omega}_{t+1}=\tilde{\varrho}\widehat{\cos(\lambda)}\tilde{\omega}_{t}+\tilde{\varrho} \widehat{\sin(\lambda)}\tilde{\omega}_{t}^\star+\tilde{\kappa}_t, \ \ \ \ \ \  \tilde{\kappa}_t\stackrel{iid}\sim \  N_m(0,\Sigma_\omega),\\
		\tilde{\omega}_{t+1}^\star=-\tilde{\varrho}\widehat{\sin(\lambda)}\tilde{\omega}_{t}+\tilde{\varrho} \widehat{\cos(\lambda)}\tilde{\omega}_{t}^\star+\tilde{\kappa}_t^\star , \ \ \ \ \  \tilde{\kappa}_t^\star\stackrel{iid}\sim \  N_m(0,\Sigma_\omega).
	\end{gathered}
\end{equation}
Here, $\tilde{\varrho}$ is a $m\times m$-dimensional diagonal matrix as a damping factor for target series $y^{(i)}$ such that $0<\varrho_{ii}<1$. $\widehat{\sin(\lambda)}$ (resp. $\widehat{\cos(\lambda)}$) is a $m\times m$-dimensional diagonal matrix whose diagonal entries equal to $\sin(\lambda_{ii})$ (resp. $\cos(\lambda_{ii})$), where $\lambda_{ii}=2\pi/q_i$ is the frequency with $q_i$ being a period such that $0<\lambda_{ii}<\pi$. 
In the cycle component, the covariance matrix for the error terms $\tilde{\kappa}_t$ and $\tilde{\kappa}_t^\star$ are assumed to be $m\times m$-dimensional diagonal matrices.

\begin{figure}[t!]
	\centering
	\subfloat[rho{[1]}$=0.06$]{\includegraphics[width = 0.47\textwidth]{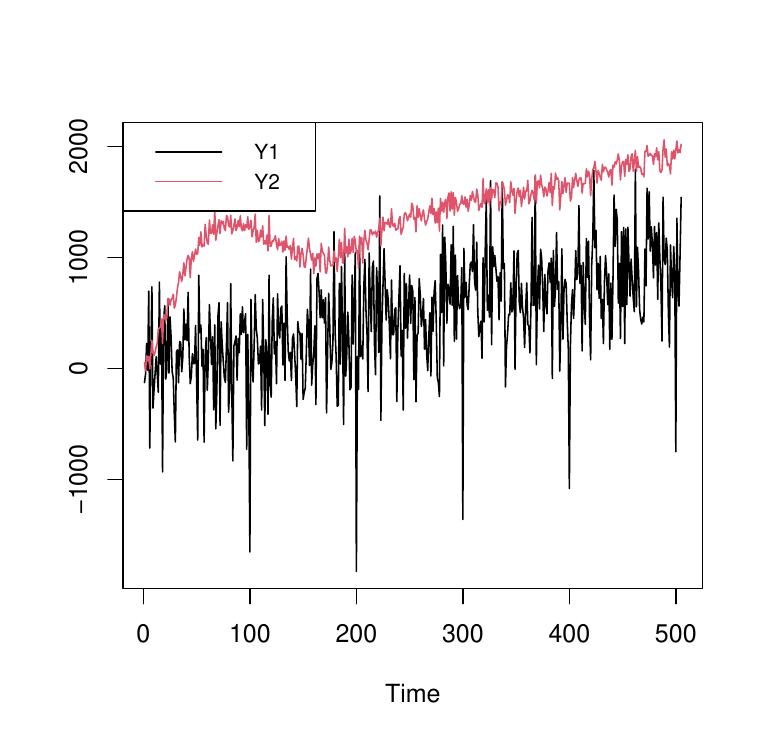}}\hfil
	\subfloat[rho{[2]}$=0.66$]{\includegraphics[width = 0.47\textwidth]{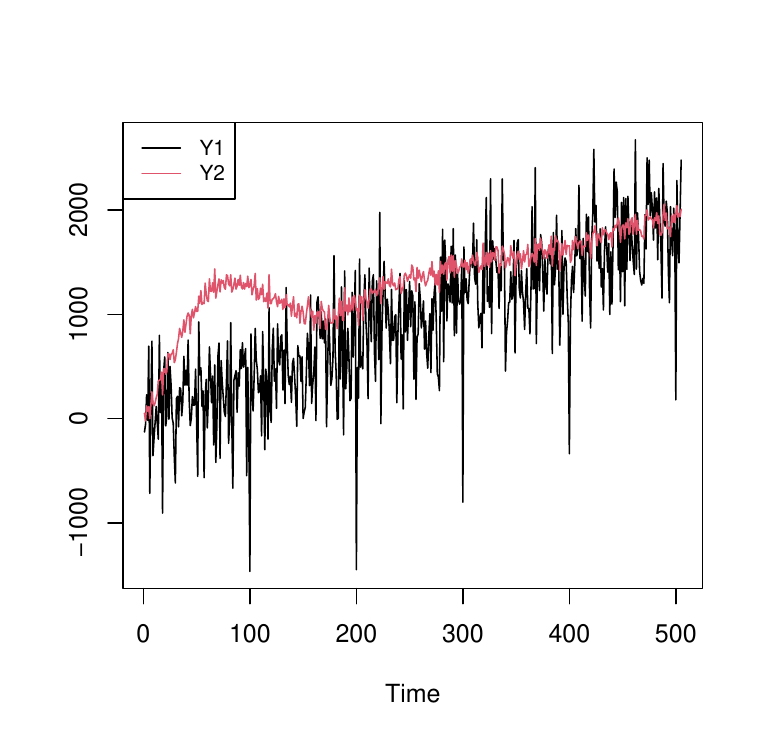}}
	\caption{Plot of two target time series.}
	\label{fig:Y12plot}
\end{figure}

In \code{data}, both $y_t^{(1)}$ and $y_t^{(2)}$ are designed to have a trend component, $\tilde{\rho}$ is $0.06$ for $y_t^{(1)}$ and $0.08$ for $y_t^{(2)}$, and $\tilde{D}$ is $-0.1$ for $y_t^{(1)}$ and $0.3$ for $y_t^{(2)}$.
In \code{data}, we set the mean of each of the diagonal entries equals to $1$ and the standard deviation of each of the diagonal entries equals to $0.5$.
In \code{data}, $y_t^{(1)}$  is generated having a seasonal component with seasonality  $S_1=100$ and $y_t^{(2)}$  is generated without a seasonal component.
In \code{data}, we set the mean of each of the diagonal entries equals to $20$ and the standard deviation of each of the diagonal entries equals to $0.5$.
In \code{data}, $y_t^{(1)}$  is generated without a cycle component and $y_t^{(2)}$  is generated having a cycle component with the damping factor being $0.99$ and the cyclic frequency being $\pi/100$.
In \code{data}, we set the mean of each of the diagonal entries equals to $20$ and the standard deviation of each of the diagonal entries equals to $0.5$. One can also set the mean and standard deviation of the entries of the covariance matrices of the state components in the \code{sim\_data} function, whose default is the following: 
\code{mean\_trend=1}, \code{sd\_trend=0.5}, \code{mean\_season=20}, \code{sd\_season=0.5}, \code{mean\_cycle=20}, \code{sd\_cycle=0.5.}
\begin{example}     
R> #Simulated data
R> data=sim_data(X=X, beta=beta, cov, k=c(8,8),  mu=c(1,1), rho=c(0.06,0.08), 
R> +            Dtilde=c(-0.1,0.3), Season=c(100,0), vrho=c(0,0.99), lambda=c(0,pi/100))
\end{example}
Now, we plot the simulated data
\begin{example}   
R> #Plot simulated data
R> ts.plot(data[,1:2],col = 1:ncol(data[,1:2]))
R> legend("topleft", colnames(data[,1:2]),col=1:ncol(data[,1:2]), lty=1, cex=1.1)
\end{example}
From Figure \ref{fig:Y12plot}, we can see that $y_t^{(1)}$ and $y_t^{(2)}$ are very different, which are generated intentionally to illustrate the strength of the MBSTS model. Clearly, $y_t^{(1)}$ is very hard to predict in that it has frequent and irregular fluctuations, as well as no sign of trend and momentum. $y_t^{(2)}$ is more stable with an obviously increasing mode. We note that most times series in real life behave more stable than $y_t^{(1)}$ and less stable than $y_t^{(2)}$, depending on the time interval and the nature of data. For example, the stock market data is much closer to $y_t^{(2)}$. We refer to \citep{qiu2018multivariate} for more numerical and empirical examples.

The two plots in Figure \ref{fig:Y12plot} are generated by calling the \code{sim\_data} function with \code{rho=c(0.06,0.08)} and \code{rho=c(0.66,0.08)}, respectively. We can see that both time series have an increasing trend, hence one could set \code{mu=c(1,1)} indicating each has a trend component. We can see that 
with a learning rate of $0.66$ the time series increases much faster, giving guidance on tuning the learning rate \code{rho}. We can see that the first time series shows a regular periodic mode every $100$ time points, which indicates that one could set the seasonality as $100$. The second time series does not show any regular mode, hence no seasonality. Instead, it has an irregular fluctuation in the range of $200$, hence one could set the cyclic frequency as $\pi/100$. Here, we observe a clear damping effect and hence the damping factor could be set as a number slightly smaller than $1$. 

\section{Preparation and the \code{tsc.setting} function}
In this section, we illustrate the prior distribution setup with the MBSTS model by providing only necessary concepts and formula to understand and use the algorithm. The R script  used in the illustration example is covered in the supplement of this paper (\code{illustration.example.R}).
\begin{example}
R> #Two target series
R> Y<-as.matrix(data[,1:m])
R> #Sixteen candidate predictors
R> X.star.single<-as.matrix(data[,(m+1):dim(data)[2]])
R> X.star<-cbind(X.star.single, X.star.single) 
R> 
R> #split dataset into training set and test set
R> n=dim(Y)[1]
R> ntrain=n-5
R> Ytrain<-Y[1:ntrain,]
R> Xtrain<-X.star[1:ntrain,]
R> Ytest<-Y[(ntrain+1):n,]
R> Xtest<-X.star[(ntrain+1):n,]
\end{example}

Generation of the initial time series components in the \textbf{mbsts} package is through the \code{tsc.setting} function. The other input parameters of the \code{tsc.setting} function are the following: The trend inclusion parameter \code{mu} and  the learning rate parameter \code{rho} for the trend component; The seasonality parameter \code{S} for the seasonal component; The damping factor parameter \code{vrho} and the frequency parameter \code{lambda} for the cycle component. 
\begin{example}
R> #Specify time series components
R> STmodel<-tsc.setting(Ytrain,mu=c(1,1),rho=c(0.06,0.08),S=c(100,0),
+                       vrho=c(0,0.99),lambda=c(0,pi/100))
\end{example}
The \code{tsc.setting} function is built upon the `SSModel' function in the state-of-the-art  package \pkg{KFAS} in \citep{helske2016kfas} and the output of the \code{tsc.setting} function is an object of `SSModel' class. 
\begin{example}
R> class(STmodel)
[1] "SSModel"
\end{example}


\section{Model Training}
Model training with the MBSTS model is performed through the \code{mbsts\_function} function in the \textbf{mbsts} package. 
The MBSTS model is the one that successfully used feature selection in multivariate time series analysis. The multivariate structure and the Bayesian framework allow the model to take advantage of the association structure among target series, and enable feature selection and model training be conducted at the same time. 
Let 
$Y=[\tilde{y}_1,\cdots,\tilde{y}_n]^T$, and we see that $Y$ is a $n\times m$ matrix. 
Then set $\tilde{Y}=\operatorname{vec}(Y)$ to transfer $Y$ to be a $nm\times 1$ vector.
Similarly, define 
$M=[\tilde{\mu}_1,\cdots,\tilde{\mu}_n]^T$,  $T=[\tilde{\tau}_1,\cdots,\tilde{\tau}_n]^T$, $W=[\tilde{\omega}_1,\cdots, \tilde{\omega}_n]^T$, and $E=[\tilde{\epsilon}_1,\cdots, \tilde{\epsilon}_n]^T$,
and set
$\tilde{M}=\operatorname{vec}(M)$, $\tilde{T}=\operatorname{vec}(T)$, $ \tilde{W}=\operatorname{vec}(W)$, $\tilde{E}=\operatorname{vec}(E)$. 
Next, we define the regression matrix $X$ and its corresponding coefficient $\beta$
\begin{equation} \label{eq:stack}
X=\begin{bmatrix}
X_1 & 0 & 0 & \dots  & 0 \\
0 & X_2 & 0 & \dots  & 0 \\
\vdots & \vdots & \vdots & \ddots & \vdots \\
0 & 0 & 0 & \dots  & X_{m}
\end{bmatrix},
\quad\quad \beta= \begin{bmatrix}
\beta_1  \\
\beta_2  \\
\vdots  \\
\beta_m
\end{bmatrix}.
\end{equation}
Here, $X$ is of dimension $(nm\times K)$ with $K=\sum_{i=1}^{m} k_i$, and its diagonal block matrix $X_i$ is a $n \times k_i$ matrix, representing all observations of  $k_i$ candidate
predictors for $y^{(i)}$. 
Now we are ready to define
$\tilde{Y}^\star=\tilde{Y}-\tilde{M}-\tilde{T}-\tilde{W}$,
and rewrite the model as
\begin{equation*} 
\tilde{Y}^\star=X\beta+\tilde{E}.
\end{equation*}

Note that, for $i=1,\cdots,m$, the $k_i$-dimensional vector $\beta_i=[\beta_{i1},\dots,\beta_{ij},\dots,\beta_{ik_i}]^T$ represents corresponding static regression coefficients for target series $y^{(i)}$, and $\beta$ defined in \eqref{eq:stack} is a $K$-dimensional vector where $K=\sum_{i=1}^{m} k_i$. Correspondingly, we define $\gamma=[\gamma_1,\dots,\gamma_m]$, where $\gamma_i=[\gamma_{i1},\dots,\gamma_{ik_i}]$ and $\gamma_{ij}=\mathbbm{1}_{\{\beta_{ij}\ne0\}}$ for $i=1,\cdots,m$ and $j=1,\cdots,k_i$. Denote $\beta_{\gamma}$ as the subset of elements of $\beta$ where $\beta_{ij}\ne 0$, and let $X_\gamma$ be the subset of columns of X where $\gamma_{ij}=1$.
A common-sense fact, if there are many predictors available then usually only a small portion of those would play a crucial role and the vast majority of regression coefficients would be zero. In the Bayesian paradigm, a natural way to represent this sparsity is through the ``spike and slab'' technique.
The ``spike'' prior is written as:
\begin{equation*} 
\gamma\sim\prod_{i=1}^{m}\prod_{j=1}^{k_i}\pi_{ij}^{\gamma_{ij}}(1-\pi_{ij})^{1-\gamma_{ij}},
\end{equation*}
where $\pi_{ij}$ is the prior inclusion probability of the $j$-th predictor for the $i$-th target time series. One can set  $\pi_i=q_i/k_i$ as the expected model size for simplicity, where $q_i$ is the number of expected nonzero predictors for $y^{(i)}$ and $k_i$ is the total number of candidate predictors for $y^{(i)}$; one can also set $\pi_{ij}$ as $0$ or $1$ to ensure certain variables to be excluded or included. In the illustration example, we set \code{ki}$=(8,16)$ as the location index of the last predictor for each target series. We set \code{pii} describe the prior inclusion probabilities $\{\pi_{ij}\}$ of each candidate predictor and initialize $\pi_{ij}=0.5$ for $i=1,\cdots,m$ and $j=1,\cdots,k_i$.
\begin{example}
R> #prior parameters setup
R> ki<- c(8,dim(Xtrain)[2])
R> pii<- matrix(rep(0.5,dim(Xtrain)[2]),nrow=dim(Xtrain)[2])
\end{example}

The coefficient $\beta$ and the covariance matrix $\Sigma_\epsilon$ are assumed to be prior independent given $\gamma$,
$$p(\beta,\Sigma_\epsilon,\gamma)=p(\beta|\gamma)p(\Sigma_\epsilon|\gamma)p(\gamma).$$
The ``slab'' prior specification is given as below, which is called ``slab'' because one can choose the prior parameters to make it only very weakly informative (close to flat), conditional on $\gamma$,
\begin{equation} \label{eq:slab}
\beta|\gamma \sim N_K(b_\gamma,(\kappa X^T_\gamma X_\gamma/n)^{-1}),\quad\quad
\Sigma_\epsilon|\gamma\sim IW(v_0,(v_0-m-1)(1-R^2)\Sigma_y).
\end{equation}
That is, $\beta$ given $\gamma$ is distributed according to the multivariate normal distribution with mean vector $b_\gamma$ and covariance matrix $(\kappa X^T_\gamma X_\gamma/n)^{-1}$, where $b_\gamma$ is the vector of prior means 
with the same dimension as $\beta_\gamma$ and $\kappa$ is the number of observations worth of weight on the prior mean vector. 
Here, $\Sigma_\epsilon$  given $\gamma$ is distributed according to the inverse Wishart distribution $IW(v_0,(v_0-m-1)(1-R^2)\Sigma_y)$, where $v_0$ is the number of degrees of freedom whose value must be greater than the dimension of $\tilde{y}_t$ plus one. In the illustration example, since there are $2$ target series, we set $v_0=5$ which is larger than $2+1$. We recommend setting $v_0$ be the dimension of $\tilde{y}_t$ plus $a\in \{2,3\}$.
We use the default that $b_\gamma=(0,0)^T$, $\kappa=0.01$ and $R^2=0.8$ (b$=$NULL, kapp $=0.01$ and R$2=0.8$ in the \code{mbsts\_function} function).

By the assumption that all components are independent of each other,  the prior distributions in multivariate form can be reduced to their univariate counterparts.
The prior distributions of the covariance matrices $\Sigma_{\mu}$ and $\Sigma_{\delta}$ for the trend component follow the inverse-gamma distribution as
$$\Sigma_{\mu}\sim IG(w_{\mu},W_{\mu}),\quad \Sigma_{\delta}\sim IG(w_{\delta},W_{\delta}).$$
The prior distribution of the covariance matrice $\Sigma_{\tau}$ for the trend component follows the inverse-gamma distribution as
$$\Sigma_{\tau}\sim IG(w_{\tau},W_{\tau}).$$
The prior distribution of the covariance matrice $\Sigma_{\omega}$ for the cycle component follows the inverse-gamma distribution as
$$\Sigma_{\omega}\sim IG(w_{\omega},W_{\omega}).$$
In the illustration example, for simplicity, we use the default value that $w_{u}=W_{u}=0.01$ for $u\in \{\mu,\delta,\tau,\omega\}$ (v $=0.01$, ss $=0.01$ in the \code{mbsts\_function} function).

Model training uses the
MCMC approach, which is to sample from a probability distribution based on constructing a Markov chain that has the desired distribution as its equilibrium distribution.  That is, the states of the chain after discarding some steps as ``burn-in'' data,
are used as samples from the desired distribution. Specifically, the MBSTS model uses the Gibbs sampler for feature selection utilizing the classical ``spike and slab'' prior setup \citep{george1997approaches}.  
Gibbs sampler can be seen as a special case of the Metropolis–Hastings algorithm. The point of the Gibbs sampler is that given a multivariate distribution it is simpler to sample from a conditional distribution than to marginalize by integrating over a joint distribution. To implement Gibbs sampler in model training, all necessary conditional probabilities were derived.  In the illustration example, we run $400$ MCMC iterations while discarding the results of the first $100$ iterations.
\begin{example}
R> #train a mbsts model
R> mbsts.model<-mbsts_function(Ytrain,Xtrain,STmodel,ki,pii,v0=5,mc=400,burn=100)
\end{example}
Each MCMC iteration runs Algorithm \ref{algo:slide_generator}, where the samples from posterior distributions of the model are generated sequentially.
\begin{algorithm}
	\caption{MBSTS Model Training \citep{qiu2018multivariate}}
	\label{algo:slide_generator}
	\begin{algorithmic}[1]
		\State Draw the latent state $\alpha=(\tilde{\mu},\tilde{\delta},\tilde{\tau},\tilde{\omega})$ from given model parameters and $\tilde{Y}$, namely $p(\alpha|\tilde{Y},\theta,\gamma,\Sigma_\epsilon,\beta)$, using the posterior simulation algorithm from \citep{durbin2002simple}.
		\State Draw time series state component parameters $\theta=(\Sigma_\mu,\Sigma_\delta,\Sigma_\tau,\Sigma_\omega)$ given $\alpha$, namely simulating
		$\theta \sim p(\theta|\tilde{Y},\alpha)$.
		\State Loop over $i$ in an random order, draw each  $\gamma_i|\gamma_{-i},\tilde{Y},\alpha,\Sigma_\epsilon$, namely simulating $\gamma \sim p(\gamma|\tilde{Y}^\star,\Sigma_\epsilon)$  one by one, using the stochastic search variable selection (SSVS) algorithm from \citep{george1997approaches}.
		\State Draw $\beta$ given $\Sigma_\epsilon$, $\gamma$, $\alpha$ and $\tilde{Y}$, namely simulating $ \beta \sim p(\beta|\Sigma_\epsilon,\gamma, \tilde{Y}^\star)$.		
		\State Draw $\Sigma_\epsilon$ given $\gamma$, $\alpha$, $\beta$ and $\tilde{Y}$, namely simulating $\Sigma_\epsilon \sim p(\Sigma_\epsilon|\gamma,\tilde{Y}^\star,\beta)$.
		
	\end{algorithmic}
\end{algorithm}

The output of the \code{mbsts\_function} function is an object of the `mbsts' class, which is defined in the \pkg{mbsts} package with slots: \code{Xtrain}, \code{Ind},
\code{beta.hat}, 
\code{B.hat},
\code{ob.sig2},
\code{States},
\code{st.sig2},
\code{ki},
\code{ntrain},
\code{mtrain}. \code{Xtrain} contains all candidate predictor series for each target series. \code{Ind} is a matrix containing MCMC draws of the indicator variable.
\code{beta.hat} is a matrix containing MCMC draws of regression coefficients. 
\code{B.hat} is an array generated by combining beta.hat for all target series. 
\code{ob.sig2} is an array containing MCMC draws of variance-covariance matrix for residuals.
\code{States} is an array containing MCMC draws of all time series components.
\code{st.sig2} is a matrix containing MCMC draws of variances for time series components. 
\code{ki} is a vector of integer values denoting the accumulated number of predictors for the target series. For example, if there are three target series where the first has $8$ predictors, the second has $6$ predictors, and the third has $10$ predictors, then the vector is c($8,14,24$).
\code{ntrain} is a numerical value for the number of observations.
\code{mtrain} is a numerical value for the number of response variables.
\begin{example}
R> class(mbsts.model)
[1] "mbsts"
attr(,"package")
[1] "mbsts"
\end{example}

\section{Training results and their associated functions}
\begin{figure*}[htbp!]
	\subfloat[]{\includegraphics[width=0.5\textwidth]{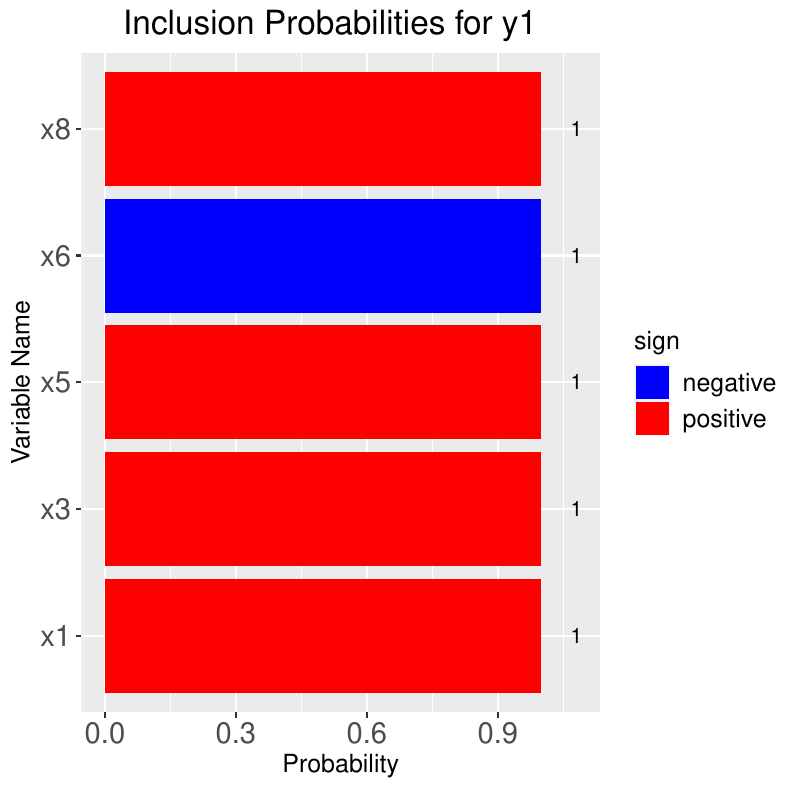}}
	\subfloat[]{\includegraphics[width=0.5\textwidth]{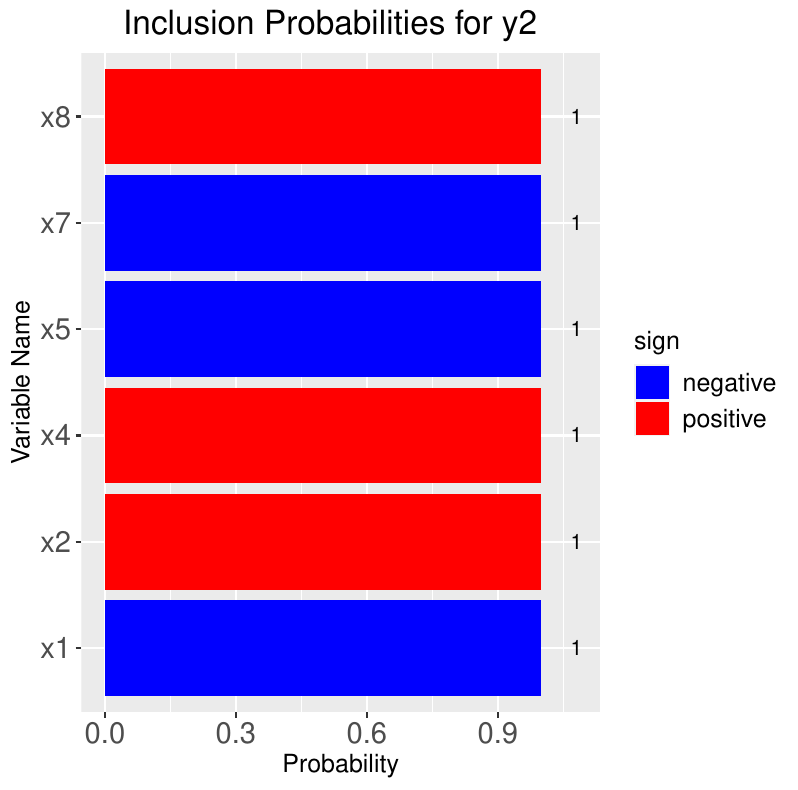}}
	\caption{Empirical posterior distribution of estimated coefficients and indicators. The red color shows positive estimated values of regression coefficients, while blue color displays negative values.}
	\label{pic:inclusion}
\end{figure*}

Recall that, in \code{data}, the regression component  is generated by 
$$\begin{bmatrix}
\beta_1 & \beta_2 & \beta_3 & \beta_4 & \beta_5 & \beta_6 & \beta_7 & \beta_8 & \\
\verteq &\verteq &\verteq &\verteq &\verteq &\verteq &\verteq &\verteq &\\
2 & 0 & 2.5 & 0 & 1.5 & -2 & 0 & 3.5
\end{bmatrix}$$
for target time series $y_t^{(1)}$, and
$$\begin{bmatrix}
\beta_1 & \beta_2 & \beta_3 & \beta_4 & \beta_5 & \beta_6 & \beta_7 & \beta_8 & \\
\verteq &\verteq &\verteq &\verteq &\verteq &\verteq &\verteq &\verteq &\\
-1.5 & 4 & 0 & 2.5  & -1 & 0 & -3 & 0.5
\end{bmatrix}$$
for target time series $y_t^{(2)}$.
 Function \code{plot\_prob} in the \textbf{mbsts} package, is developed to retrieve information from an object of the `mbsts' class to generate plots for empirical posterior distributions of estimated coefficients and indicators, for each target time series. 
 The empirical posterior inclusion probability, as a useful indicator of the importance of one specific predictor, is the proportion of the number of times that the predictor is selected to the total count of MCMC iterations after discarding the ``burn-in'' iterations.
 Users can set their desired threshold value of inclusion probability through \code{prob.threshold},  rename the predictors through \code{varnames}, and set titles through \code{title} for each plot. The default threshold value of inclusion probability is $0.8$.
\begin{example}
R> #title vector for each plot
R> title_new<-c("Inclusion Probabilities for y1", 
+	        "Inclusion Probabilities for y2")
	
R> #rename predictors
R> varnames_new<-c("x1", "x2", "x3", "x4", "x5", "x6", "x7", "x8", 
+	          "x1", "x2", "x3", "x4", "x5", "x6", "x7", "x8")
	
R> #plot inclusion probability
R> plot_prob(object=mbsts.model,title=title_new,prob.threshold=0.8,
+            varnames=varnames_new)
\end{example}
Figure \ref{pic:inclusion} provides the empirical posterior distribution of estimated indicators and signs of corresponding coefficients. 
The probability $1$ of a specific feature indicates that it was selected in every MCMC iteration after discarding the ``burn-in'' iterations. 
We can see that the features with non-zero coefficients all were selected with correct signs indicated. 

The \code{para.est} function is developed to retrieve information from an object of the `mbsts' class to provide 
parameter estimation results for selected predictors. The output is a list: \code{\$index} provides the index of selected predictors according to the threshold value $0.8$, \code{\$para.est.mean} provides the estimated parameter values for those selected predictors, \code{\$para.est.sd} provides the standard deviations for parameter estimations.
\begin{example}
R> #Generate feature selection and parameter estimation results
R> para.est(object=mbsts.model,prob.threshold=0.8)
$index
[1]  1  3  5  6  8  9 10 12 13 15 16

$para.est.mean
[1] 1.9907  2.4501  1.4931 -1.9473  3.4812 -1.5023  3.9914  2.5143 -1.0007 -2.9951  0.4977

$para.est.sd
[1] 0.0321 0.3245 0.0319 0.1714 0.0072 0.0063 0.0541 0.0300 0.0077 0.0417 0.0018
\end{example}

\begin{figure*}
	\centering
\includegraphics[height=6.5cm, width=0.75\textwidth]{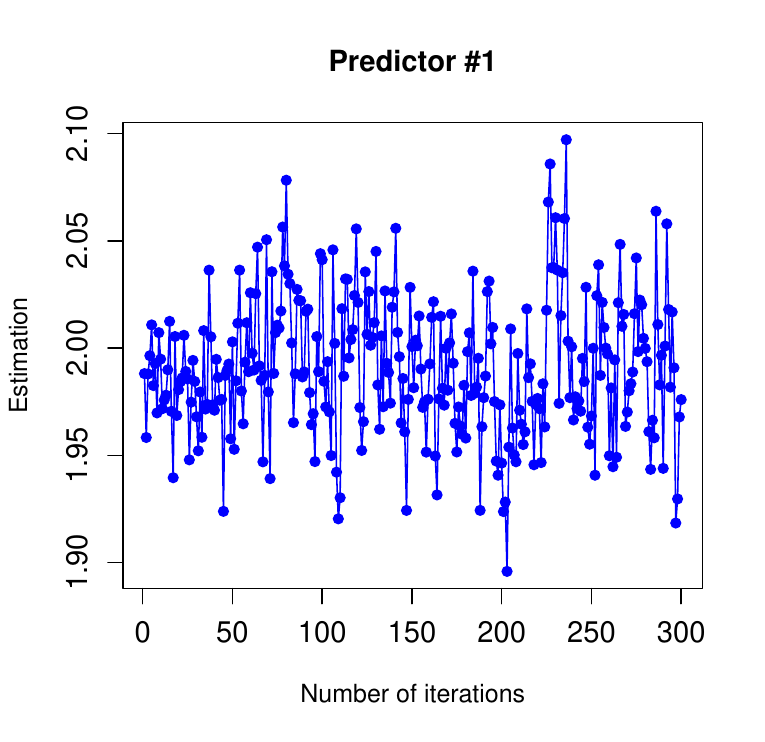}
	\caption{Convergence diagnosis for the first predicator  of $400$ MCMC iterations after discarding the first $100$.}
	\label{pic:cvg}
\end{figure*}

We conduct convergence diagnosis by calling the \code{plot\_cvg} function, using the index number of one specific parameter. Figure \ref{pic:cvg} reports the estimated values for the first parameter of MCMC iterations after ``burn-in''. 
\begin{example}
R> #Convergence diagnosis
R> plot_cvg(object=mbsts.model,index=1,main="Predictor #1")
\end{example}

\begin{figure*}
	\subfloat[]{\includegraphics[width=0.5\textwidth]{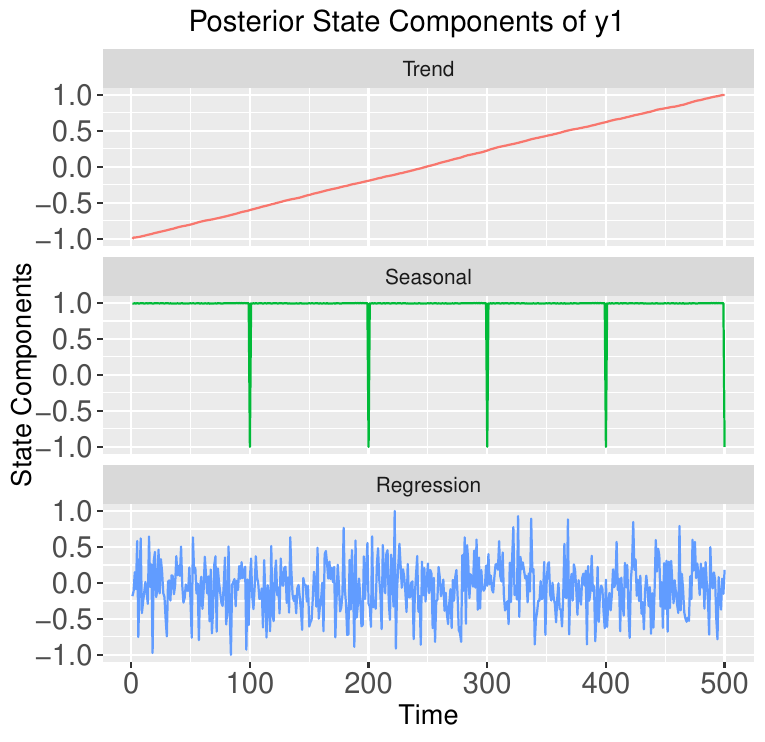}}
	\subfloat[]{\includegraphics[width=0.5\textwidth]{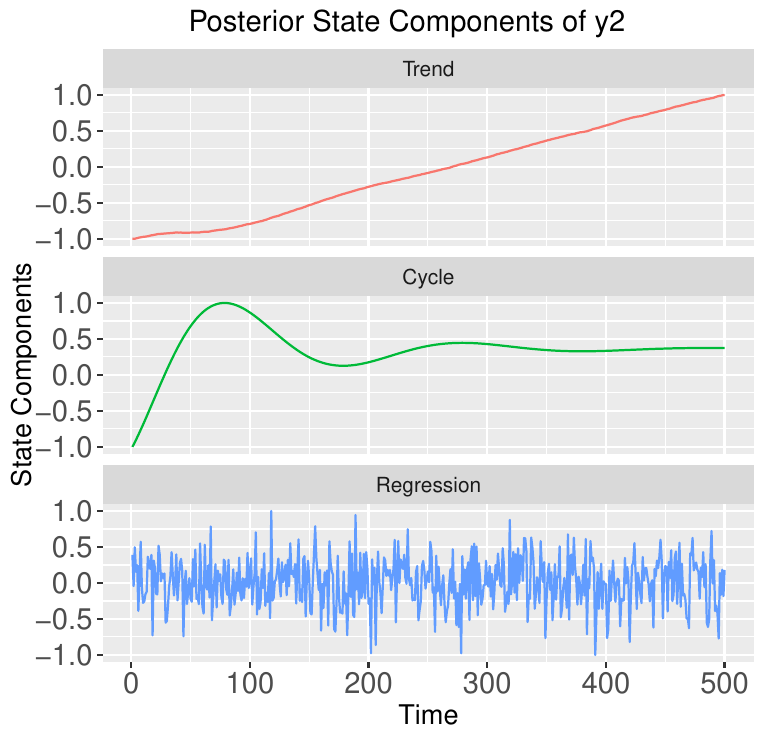}}
	\caption{Empirical posterior distributions of all estimated state components}
	\label{pic:states}
\end{figure*}

\begin{figure*}
	\subfloat[]{\includegraphics[width=0.5\textwidth]{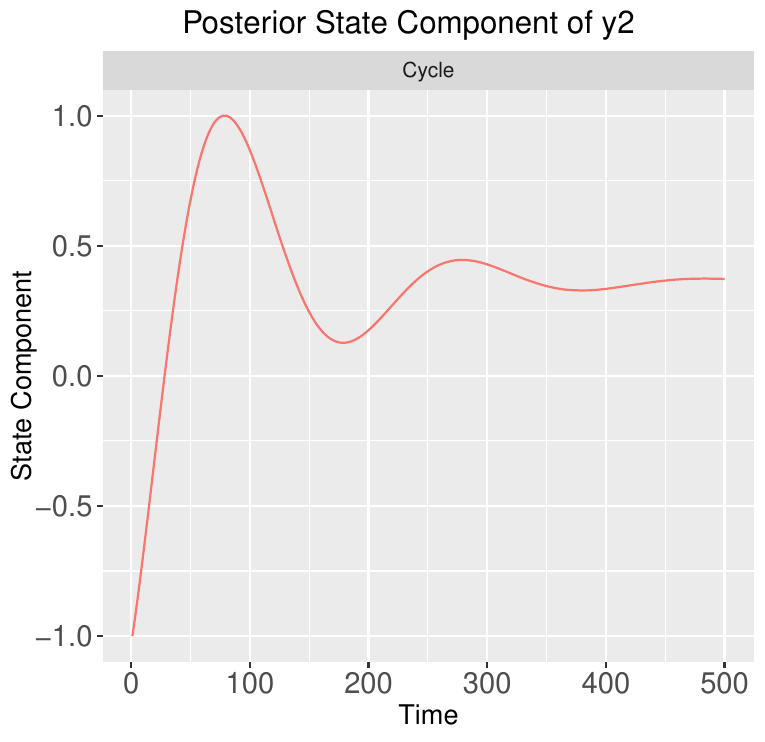}}
	\subfloat[]{\includegraphics[width=0.5\textwidth]{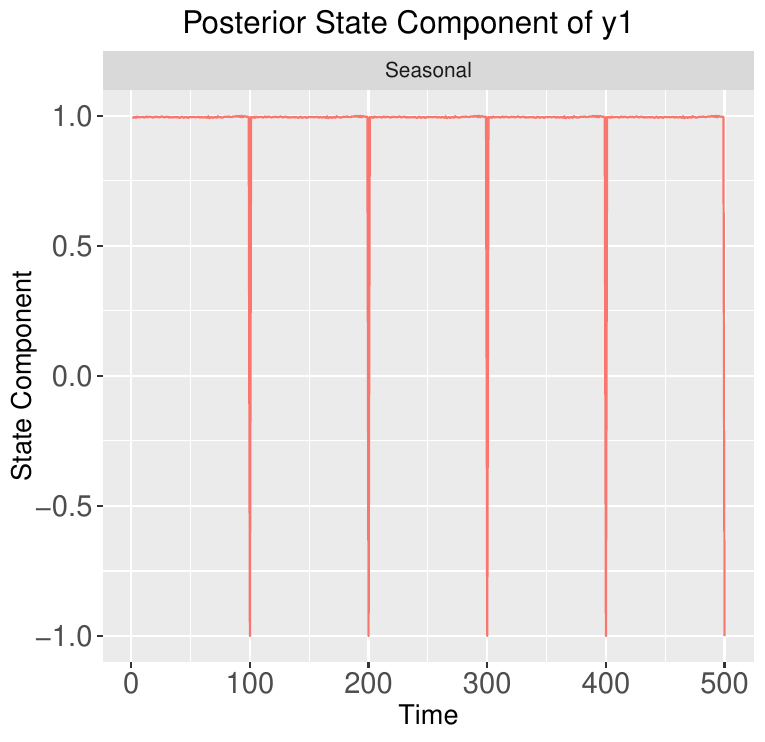}}
	\caption{Empirical posterior distributions of specific estimated state component}
	\label{pic:states2}
\end{figure*}

The \code{plot\_comp} function is developed to retrieve information from an object of the `mbsts' class to plot empirical posterior distributions of estimated state components.
The \code{plot\_comp} function in the \textbf{mbsts} package, provides the mean of empirical posterior distribution for each estimated state component. One can set \code{component\_selection="All"} to generate plots for all state components as illustrated in Figure \ref{pic:states}. One can set \code{component\_selection} to an individual state component such as \code{component\_selection="Cycle"},
to generate plots for the cycle component as illustrated in Figure \ref{pic:states2}. 
\begin{example}
R> #plot state components
R> #title vector for each plot
R> title_all<-c("Posterior State Components of y1", "Posterior State Components of 
                 y2")
R> #all components
R> plot_comp(object=mbsts.model, slope=c(T,T),local=c(T,T),season=c(100,0),
             cyc=c(F,T),title=title_all,component_selection="All")
R> #title vector for each plot
R> title_single<-c("Posterior State Component of y1", "Posterior State Component of 
                    y2")
R> #individual components
R> plot_comp(object=mbsts.model, slope=c(T,T),local=c(T,T),season=c(100,0),
             cyc=c(F,T),title=title_single,component_selection="Cycle")
R> plot_comp(object=mbsts.model, slope=c(T,T),local=c(T,T),season=c(100,0),
             cyc=c(F,T),title=title_single,component_selection="Seasonal")
\end{example}

\section{Forecasting}
\label{sec:Forecasting}
Denote $\tilde{\psi}=(\alpha,\theta,\gamma,\Sigma_\epsilon, \beta)$.
Looping through the five steps in Algorithm \ref{algo:slide_generator} yields a sequence of draws $\tilde{\psi}$ from a Markov chain with stationary distribution $p(\tilde{\psi}|Y)$ which is the posterior distribution of $\tilde{\psi}$ given $Y$. 
After model training, forecasts are based on the posterior predictive distribution. 
Let $\hat{Y}$ represent the set of values to be forecast. 
Samples of $\hat{Y}$ from $p(\hat{Y}|\tilde{\psi})$ can be drawn by simply iterating equations \eqref{eq:trend}, \eqref{eq:slope}, \eqref{eq:seasonal}, \eqref{eq:cycle}, and \eqref{eq:regression} to move forward using initial values of states $\alpha$ and initial values of parameters $\theta$, $\beta$, and $\Sigma_\epsilon$.
For example, for the one-step-ahead forecast, samples were drawn from the multivariate normal distribution with mean equal to $\tilde{\mu}_n+\tilde{\tau}_n+\tilde{\omega}_n+\tilde{\xi}_n$
and variance equal to $\Sigma_{\mu}+\Sigma_{\tau}+\Sigma_\omega+\Sigma_\epsilon$. The detailed procedure can be seen in 
Algorithm \ref{algo:forecast} where the point prediction values could be formed by taking the average of drawn samples at the end. 
\begin{algorithm}
	\caption{Model Forecast \citep{qiu2020multivariate}}
	\label{algo:forecast}
	\begin{algorithmic}[1]
		\State Draw the next latent time series states $\alpha_{t+1}=(\tilde{\mu}_{t+1},\tilde{\delta}_{t+1},\tilde{\tau}_{t+1},\tilde{\omega}_{t+1})$ given current latent time series states $\alpha_{t}=(\tilde{\mu}_{t},\tilde{\delta}_{t},\tilde{\tau}_{t},\tilde{\omega}_{t})$ and component parameters $\theta=(\Sigma_\mu,\Sigma_\delta,\Sigma_\tau,\Sigma_\omega)$, based on equations \eqref{eq:trend}, \eqref{eq:slope}, \eqref{eq:seasonal} and \eqref{eq:cycle}.
		\State Based on indicator variable $\gamma$, compute the regression component given the information about predictors at time $t+1$ by equation \eqref{eq:regression}.
		\State Draw a random error in multivariate normal distribution with variance equal to $\Sigma_\epsilon$ and sum them up using equation \eqref{eq:st}.
		\State Sum up all the predictions and divide by the total number of MCMC iterations to generate the point prediction.
	\end{algorithmic}
\end{algorithm}

The \code{mbsts.forecast} function is developed to retrieve information from an object of the `mbsts' class
and an object of the `SSModel' class to forecast. 
It forecasts multiple steps ahead in the way that once a forecast is produced it is added to the training data as a ``fake'' data. In the following example, a $5$-steps-ahead forecast  is generated by setting \code{steps} in the \code{mbsts.forecast} function. 
\begin{example}
R> #make a 5-steps prediction
R> output<-mbsts.forecast(mbsts.model,STmodel,newdata=Xtest,steps=5)
\end{example}
The \code{mbsts.forecast} function has two outputs: 
\code{pred.dist} is an array of draws from the posterior distribution for the forecasts, i.e., for each time series it contains a matrix whose row indicates the MCMC iterations after ``burn-in'' and whose column indicates the steps of forecast; 
\code{pred.mean} is a matrix giving the posterior mean of the prediction for each target series.
\begin{example}
R> output$pred.mean
       [,1]     [,2]
[1,] 1316.5353 2019.864
[2,]  698.9187 1905.985
[3,]  525.5992 1937.280
[4,]  987.0031 1904.776
[5,] 1495.9308 1977.514
\end{example}

\section{Conclusion}
This paper demonstrates how to use the R package \pkg{mbsts} for MBSTS modeling. By the multivariate nature of the MBSTS model, the correlations among multiple target series are naturally taken into account, which helps boost the forecasting power. Figure 6 in \cite{qiu2018multivariate} revealed that the higher correlation among multiple target time series, the better performance of the MBSTS model over the univariate BSTS model. Therefore, it is better to model multiple target time series as a whole by MBSTS rather than model them individually by BSTS, especially when strong correlations appear in the multiple target time series. In this paper, we focus on the MBSTS modeling which uses the Bayesian paradigm to reduce dimension, but the frequentist approach could be better under different scenarios (see, e.g., \cite{ning2021scalable} in Monte Carlo setting). 


\section*{Acknowledgments}
The research of Ning Ning was partially supported by the Seed Fund Grant Award at Texas A\&M University.

\bibliography{RJreferences}
\address{Ning Ning\\
	Department of Statistics\\
	Texas A\&M University\\
	\email{patning@tamu.edu}}

\address{Jinwen Qiu\\
  Department of Statistics and Applied Probability\\
  University of California, Santa Barbara\\
  \email{qjwsnow\_ctw@hotmail.com}}

\end{article}

\end{document}